\documentclass{article}

\usepackage{arxiv}

\usepackage{amsthm}
\usepackage[utf8]{inputenc} 
\usepackage[T1]{fontenc}    
\usepackage{hyperref}       
\usepackage{url}            
\usepackage{booktabs}       
\usepackage{amsfonts}       
\usepackage{nicefrac}       
\usepackage{microtype}      
\usepackage{lipsum}		
\usepackage{graphicx}
\usepackage{doi}

\title{On computations with Double Schubert Automaton and stable maps of Multivariate Cryptography}

\author{ \href{https://orcid.org/0000-0002-2138-2357}{\includegraphics[scale=0.06]{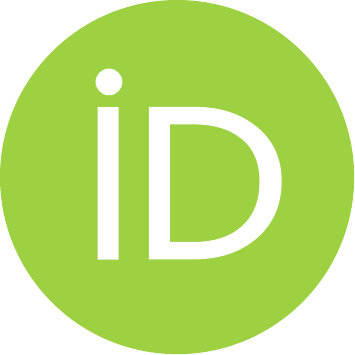}\hspace{1mm}Vasyl Ustimenko} \\
	Faculty of Mathematics, Physics and Computer Science\\
	Maria Curie-Sk{\l}odowska University\\
	Pl. Maria Curie Sk{\l}odowska 1,  20-031 Lublin, Poland\\\
	\texttt{vasyl@hektor.umcs.lublin.pl}\\
}

\renewcommand{\shorttitle}{On computations with Double Schubert Automaton and stable maps}

\hypersetup{
pdftitle={On computations with Double Schubert Automaton and stable maps of multivariate cryptography},
pdfsubject={math.RA, math.GR, cs.CR},
pdfauthor={Vasyl Ustimenko},
pdfkeywords={Affine Cremona Group, Double Schubert Graphs, Multivariate Cryptography, Noncommutative Cryptography, Post Quantum Cryptography},
}

\begin{document}
\maketitle

\begin{abstract}
 The families of bijective transformations $G_n$ of affine space $K^n$ over general commutative ring $K$ of increasing order with the
property of stability will be constructed. Stability means that maximal degree of  elements of 
cyclic subgroup generated by the transformation of degree $d$ is bounded by $d$. 
In the case $K=F_q$ these transformations of
 $K^n$ can be of an exponential order. 
We introduce large groups formed by quadratic transformations
and numerical encryption algorithm protected by secure protocol of Noncommutative Cryptography. The construction of transformations is presented in terms of walks on Double Schubert Graphs.
\end{abstract}


\section{Introduction}

    In $2017$ the international tender of the National Institute of Standartisation Technology (NIST) of the  USA for the selection of public key based on postquantum algorithms was announced.  It has been considering algorithms for the encryption task and for the procedure of digital signature.
     The last third round of this competition started in summer time of $2020$. Only one  candidate from the multivariare cryptography area remains. This is  a special case of ‘’Rainbow like unbalanced oil and vinegar’’ digital scheme.The final list does not contain algorithms of Multivariate Cryptography for the encryption task.
      This outcome stimulates alternative research on numerical encryption  asymmetrical postquantum algorithms of Multivariate cryptography such as algorithms which are not public keys and use the composition  of several nonlinear maps of bounded degree. Our paper is dedicated to new  postquantum secure cryptosystem with the encryption process based on bijective quadratic maps of large order.
	 Postquantum status of these encryption is justified by recent results of Noncommutative Cryptography. 
	
	In March 2021 it was announced that prestigious Abel prize will be shared by A. Wigderson an L. Lovasz. They contribute valuable applications of theory of Extremal graphs (see \cite{20}) and Expanding graphs \cite{21} to Theoretical Computer Science. We have been working on applications of these graphs to Cryptography 
	(see \cite{22}, \cite{23}, \cite{24}, \cite{25} and further references). This paper is dedicated to the  problem of postquantum secure encryption of rather large files in terms of Multivariate Cryptography but with usage of ideas of Noncommutative Cryptography. We will use Double Schobert graphs which belong to class of geometrical expanders introduced in \cite{26}.
	Remarkable symbiotic combination of absolutely secure one time pad with Diffie-Hellman protocol in terms of groups $F^*_p$, $p$ is prime,
	can not be used in our postquantum times  because classical discrete logarithm problem can be solved in polynomial time vith usage of  quantum computer. The proof  of this fact waspublished by Peter Shor in 1995.
	We present a possible substitutor of mentioned above symbiotic combination.
	
	Classical encryption tools of Multivariate Cryptography are nonlinear polynomial maps $F$
of affine space $K^n$ over finite commutative ring $K$ into itself. Traditionally a map $F$ is presented in 
the form
$T_1GT_2$, where $T_1$ and $T_2$ are representatives of affine general group $AGL_n(K)$ of all polynomial bijective transformations of
$K^n$ of degree 1 and central $G$ is a nonlinear polynomial map. We refer to $F$ as linear deformation of $G$.
Popular computer tools  for the generation of $G$ are packages for symbolic computations (''Mathematica'', ''Maple'', ''Sage'',
''Magma'' and special symbolic tools for professionals).  Alternative approach to the construction of core maps $G$ via {\em numerical}
computations with sparse algebraic graphs was presented at some talks at CANA conference of FedSCIS \cite{16}, \cite{17}, \cite{18}. The idea is to convert albebraic graphs into special automata for computations in polynomial ring $K[x_1, x_2, \dots, x_n]$ in terms of ''arithmetical''
operations of addition and multiplication in the ring. It allows to use standard $C^{++}$ or Java languages for the construction of
polynomial maps over finite fields, arithmetical and Boolean rings.  It is interesting that automata was constructed from
bipartite algebraic graphs defined by systems of equations $x_i-y_i=x_ly_k$, some properties of graphs (stability, degree, in particular) were obtained theoretically but other properties (orders, density) were investigated via computer simulation.

This paper is a theoretical one, we present theoretical results which demonstrate potential of the graph based approach.
It turns out that the method allows to generate a stable nonlinear polynomial maps of chosen degree with a prescribed density and
exponentially growing order. Results are obtained via explicit constructions of automaton maps based on bipartite graph
$DS(n,K)$ over general commutative ring $K$ such that point $(x_1, x_2, \dots, x_n, x_{11}, x_{12}, \dots, x_{nn})$ is incident to line
 $[x_1, x_2, \dots, x_n, x_{11}, x_{12}, \dots, x_{nn}]$ if and only if $x_{ij}-y_{ij}=x_iy_j$, $i=1,2, \dots, n$,
$j=1,2, \dots, n$.  Special walks on this graph of even length induce
nonlinear map of affine space $K^{n(n+1)}$ to itself. The graph has geometrical nature, in case of $K=F_q$ it is induced subgraph of the
incidence relation of finite projective geometry.

 The approach was motivated by cryptographical applications. That is why explicit constructions lead to some new cryptosystems.
In Section 2 we discuss the concepts of postquantum security and multivariate cryptography (MC), some references on usage of cryptographical properties of algebraic graphs are given. Next section is devoted to the concept of stable transformation connected with the investigation
of discrete logarithm problem in the affine Cremona group $C(K^n)$ of all bijective polynomial transformations of affine space $K^n$ such that their inverses are also polynomial maps. This problem is motivated by related  multivariate Diffie - Hellman key exchange protocol and corresponding El Gamal cryptosystem. In Section 4 we modify El Gamal algorithm, one can use high non commutativity of $C(K^n)$ and conjugate the inverse of the generator of large cyclic group. In the next section we state theorems on the existence of families of nonlinear
stable multivariate maps over finite fields of exponentially growing order with prescribed degree and density. The existence of corresponding
explicit construction is also formulated.  The impact of such theorems is an option of implementation
of multivariate key exchange protocols and related shifted El Gamal cryptosystems in case of family of cyclic subgroups of exponentially growing order in affine Cremona group over $F_q$. In Section 7 we discuss natural restriction on parameters for such algorithm.
The Double Schubert graph and related automaton are introduced in Section 8. The sketch of proof of the main theorem on the existence of
 stable maps
of exponentially growing order is given as a chain of lemmas.

The last section of the paper present new encryption algorithm of Post Quantum Cryptography. Encryption is descrribed in terms of 
Quadratic Multivariate Cryptography. The speed of encryption is standard for this area. Suggested cryptosystem is Post Quantum Secure.
It is not a public key. So it differs from candidates investigated by well known NIST competition.
Cryptosystem combines secure protocol with quadraticsable platform of polynomial transformations with flexible encryption procedure.


\section{On Post Quantum and Multivariate Cryptography}

    Post Quantum Cryptography serves for the research of asymmetrical
  cryptographical algorithms which can be potentially resistant
  against attacks based on the use of quantum computer.

  The security of currently popular algorithms is based on the complexity of the
  following three well known hard problems: integer factorisation, discrete logarithm problem,
discrete logarithm for elliptic curves.
Each of these problems can be solved in polynomial time by Peter
    Shor's algorithm for theoretical quantum computer.
So cryptographers already started
    research on postquantum security.
        They also have to investigate the impact of the new results on general complexity theory such as
        complexity estimates of graph isomorphism problem obtained by L. Babai \cite{11}.


       We have to notice that Post Quantum Cryptography (PQC) differs from
       Quantum Cryptography, which is based on the idea of usage of
       quantum phenomena to reach better security.

       Modern PQC is divided into several directions such as
       Multivariate Cryptography, Lattice based Cryptography, Hash
       based Cryptography, Code based Cryptography, studies of
       isogenies for superelliptic curves.

The oldest direction is Multivariate Cryptography which uses polynomial maps of affine
space $K^n$ defined over a finite commutative ring into itself as
an encryption tool. It exploits the complexity of finding
solution of a system of nonlinear equations from many variables.

This is still young promising research area with the current lack of
known cryptosystems with the proven resistance against attacks with
the use of ordinary Turing machines. Studies of attacks based on Turing machine and Quantum computer
have to be investigated separately because of different nature of two machines, deterministic and probabilistic
respectively.
 Multivariate cryptography started from the studies of
potential for the special quadratic encryption multivariate
bijective map of ${K}^n$, where $K$ is an extention of finite field
$F_q$ of characteristic 2. One of the first such cryptosystems was
proposed by Imai and Matsumoto, cryptanalysis for this system was
invented by J. Patarin. The survey on various modifications of this
algorithm and corresponding cryptanalysis the reader can find in
\cite{1}. Various attempts to build secure multivariate public key were
 unsuccessful, but the research of the development of new candidates for secure multivariate
public keys is going on (see for instance \cite{2} and further references).

  Applications of Algebraic Graph Theory to Multivariate Cryptography were recently observed in \cite{3}.
  This survey is devoted to
  algorithms based on bijective maps of affine spaces into itself.

\section{On stable multivariate transformations for the key exchange protocols}

It is widely known that  Diffie - Hellman key exchange protocol can be formally considered for the generator $g$ of a finite group or
 semigroup $G$.
Users need a large set $\{g^k|k=1,2, \dots \}$ to make it practical. One can see that security of the method depends not only on abstract
group or semigroup $G$ but on the way of its representation.  If $G$ is a multiplicative group $F_p^{*}$ of finite field $F_p$ than we have a case of classical Diffie - Hellman algorithm. If we change ${F_p}^{*}$
for isomorphic group $Z_{p-1}$ then the security will be completely lost. We get a problem of solving linear equation instead of a
discrete logarithm problem to measure the security level.

Let $K$ be a commutative ring. $S(K^n)$ stands for the affine Cremona semigroup of all bijective polynomial transformations of affine space 
$K^n$.

Let us consider a multivariate Diffie - Hellman key exchange algorithm for the generator $g(n)$ semigroup $G_n$ of affine Cremona semigroup.
Correspondents (Alice and Bob) agree on the generator $g(n)$ of  group  of kind

$x_1 \rightarrow f_1(x_1, x_2, \dots, x_{n})$,  $x_2 \rightarrow f_2(x_1, x_2, \dots, x_{n}), \dots$,  $x_{n} \rightarrow f_{n}(x_1, x_2, \dots, x_{n})$
acting on the affine space $K^n$, where $f_i \in K[x_1, x_2, \dots, x_{n}]$,
$i=1, 2, \dots,  n$ are multivariate polynomials.
Alice chooses  a positive integer $k_A$ as her private key and computes the transformation ${g(n)}^{k_A}$
(multiple iteration of $g(n)$ with itself).

Similarly Bob chooses $k_B$ and gets ${g(n)}^{k_B}$. Correspondents complete the
exchange: Alice sends ${g(n)}^{k_A}$ to Bob and receives  ${g(n)}^{k_B}$ from him.
 Now Alice and Bob computes independently common key $h$ as $({g(n)}^{k_B})^{k_A}$ and $({g(n)}^{k_A})^{k_B}$ respectively.
So they can use coefficients of multivariate map $h={g(n)}^{{k_B}{k_A}}$ from $G_n$ written in the standard form.

There are obvious problems preventing the implementation of this general method in practice.
In case $n=1$ the degree ${\rm deg}(fg)$ of composition $fg$ of elements $f, g \in S(K)$ is simply the product
of ${\rm deg}(f)$ and ${\rm deg}(g)$. Such effect can happen in multidimensional case:
$({\rm deg} (g))^x={\rm deg} (g^x)=b$. It causes the reduction of discrete logarithm problem for multivariate polynomials
to a number theoretical problem. If $g$ is a bijection of degree $d$ and order $m$ then $d^x=b$ in cyclic group $Z_m$.
Similar reduction can appear in the case of other degree functions $s(x)={\rm deg}(g^x)$. If $s(x)$ is a linear function then multivariate
discrete logarithm problem with base $g$ is reducible to the solution of linear equation.
The degenerate case ${\rm deg}(g^x)={\rm const}$ is an interesting one because in such situation the degree function does not help to
investigate multivariate discrete logarithm.

We refer to the sequence of multivariate transformations $f(n) \in S(K^n)$ as stable maps of degree $d$ if
 ${\rm deg}(f(n))$ is a constant $d, d>2$ and  ${\rm deg}(f(n)^k)\le d$ for $k=1, 2, \dots$ (see \cite{3}).
If $\tau_n$ is a bijective affine transformation of $K^n$, i. e. a bijective transformation of degree 1,
then the sequence of stable maps $f(n)$ can be changed for other sequence of stable maps $\tau f(n){\tau}^{-1}$ of the same degree $d$.

The first families of special bijective transformations of $K^n$ of
bounded degree were generated by {\em discrete dynamical systems}
defined in \cite{4} in terms of graphs $D(n, K)$. In the paper \cite{5} the
fact that each transformation from these families of maps is cubic
was proven. In \cite{6} authors notice that this family is a stable one, the
order of its members grows with the increase of parameter $n$ and
suggests key exchange protocols with generators from these families.
Other results on the usage of algebraic graphs for construction of families of nonlinear multivariate maps of
 degree $\le 3$  the reader can find in \cite{7}, \cite{8}.



 Recall that the other important property for the generator $g(n)$ in the described above protocol is a 
large cardinality
of  $\{g(n)^k|k=1,2, \dots \}$. Let us assume that $g(n)$ is bijection.

The famous family of linear bijections
of ${F_q}^n$ of exponential order is formed by Singer cycles $s(n)$, they have order $q^n-1$ (see \cite{11}, \cite{12} and further references). Statements on the existence of explicit construction of families of  nonlinear maps of
exponential order are formulated in the section 5 of this paper.


The above mentioned key exchange protocol can be used for the design of the
following multivariate El Gamal cryptosystem (see \cite{9}, \cite{10}).

 Alice takes generator $g(n)$ of the group $G_n$ together with its inverse $g(n)^{-1}$. She sends the transformation
$g(n)^{-1}$ to Bob. He will work with the plainspace ${K}^n$ as public user.

At the beginning of each session Alice chooses her private key $k_A$. She
computes $f=g(n)^{k_A}$ and sends it to Bob.

Bob writes his text $(p_1, p_2, \dots, p_{n})$, chooses his private key $k_B$ and
creates the ciphertext $f^{k_B}((p_1, p_2, \dots, p_{n}))={\rm c}$.

Additionally he computes the map $g(n)^{-1}{k_B}=h(n)$ and sends the pair $(c_1, c_2, \dots, c_{n}), h(n)({\rm x})$ to Alice.

  Alice computes ${h(n)}^{k_A}({\rm c})=(p_1, p_2, \dots, p_n)$.

    REMARK 1. {\em It is proven (see [9]) that the security level of above multivariate Diffie - Hellman and El Gamal algorithms is the
		same. It is based on the multivariate discrete logarithm problem on
solving the equation $g^x=d$, where $g$ and $d$ are elements of special cyclic subgroup $G_n$ of affine Cremona group}.

\section{On the shifted multivariate El Gamal cryptosystem}

We suggest here the following modification of above described algorithm.
 Alice takes generator $g(n)$ of the group $G_n$ together with its inverse $g(n)^{-1}$. 
At the beginning of each session Alice chooses her private key $k_A$  and pair $(h(n), {h(n)}^{-1})$, where $h(n)$ is
an element of affine Cremona group.
She computes $f=g(n)^{k_A}$ and sends it to Bob together with transformation $m(n)=h(n){g(n)}^{-1}{h(n)}^{-1}$.
Public user
Bob writes his text $(p_1, p_2, \dots, p_{n})$, chooses his private key $k_B$ and
creates the ciphertext $f^{k_B}((p_1, p_2, \dots, p_{n}))={\rm c}$.

Additionally he computes the map $m(n)^{k_B}=a(n)$ and sends the pair $(c_1, c_2, \dots, c_{n}), a(n)({\rm x})$ to Alice.

  Alice computes  $h(n)^{-1}{a(n)}^{k_A}h(n)({\rm c})=(p_1, p_2, \dots, p_n)$.

The shifted algorithm can have better protection against attacks by adversary. One can choose $h(n)$ to make the discrete logarithm problem
in affine Cremona group with base $m(n)$ harder than one in a case of base ${g(n)}^{-1}$.
Additionally the adversary has to compute the inverse of $f={g(n)}^{k_A}$.

Alice can work with a stable map  $g(n)$ of a large polynomial degree and a polynomial density of a large order such
that its inverse conjugate with stable map $m(n)$ of prescribed small constant degree $d$.

 REMARK 2. {\em It is clear, that the algorithm above can be formally considered for the general pair of bijective nonlinear polynomial
    transformations $g(n)$ and  $h(n)$ of affine Cremona group of the free module ${K}^n$. But the best computational complexity  will be achieved in the case of
    quadratic stable elements $g(n)$ and $m(n)={h(n)}g(n)^{-1}h(n)^{-1}$. In the case of a family $m(n)$ of exponential order corresponding discrete logarithm problem looks as a hard one}.

\section{Results on existence of families with prescribed properties} 

Recall that the density of multivariate polynomial $f\in K[x_1, x_2, \dots, x_n]$ is its number of monomial terms.

 The density of a transformation $F$ of $K^n$ given by rules 
$x_1 \rightarrow f_1(x_1, x_2, \dots, x_n)$, $x_2 \rightarrow f_2(x_1, x_2, \dots, x_n)$, $\dots$,
$x_n \rightarrow f_n(x_1, x_2, \dots, x_n)$ is defined as a maximum of densities of $f_i$, $i=1,2, \dots, n$.

We refer to $F(n): K^n \rightarrow K^n$ as a family of density $d$ if a density of $F(n)$ is estimated by $Cn^d$, where
$C$ is a positive constant. If each transformation $F(n)$ of a density $d$ has constant degree $t$, then $d \le t$.

We refer to a family of bijective linear transformation $\tau(n)$ given by rule
$(x_1, x_2, \dots, x_n) \rightarrow (x_1, x_2, \dots, x_n)A$ of affine space $K^n$ as sparse
transformations if each row and column of matrix $A(n)$ contains only finite number of nonzero entries and this number is bounded by some positive constant.  We refer to $G(n) = \tau(n)F(n)\tau(n)^{-1}$ as a sparse deformation of a family $F(n)$.
if one of the families $F(n)$ and $G(n)$ has density $d$ then the density of the other is also $d$. 

THEOREM 1 

{\em For each pair $(d, T)$, where $d \le T$ there is a family $F(n)$ of stable transformations of $K^n$, $n=k(k+1)$ of degree $T$ and density $d$ of 
order bounded below by $\sqrt{n}-1$}. 


In the case of finite fields we get the following statement.

THEOREM 2.

{\em Let $F_q$ be a finite field. For each pair $(d, T)$, where $1/2 \le d \le T$ there is a family $F(n)$ of stable transformations of $K^n$ of degree $t$, density $d$ and order at least $q^{\sqrt{n}-1}-1$}.

Sketches of proofs ot the theorems 1 and 2 are presented in the section 6. This technique is used also for the proof of the following statement.

THEOREM 3.

{\em Let $F_q$ be a finite field. For each pair $(d, T)$, where $1/2 \le d \le T$ there is a family $F(n)$ of stable transformations of ${F_q}^n$,
$n=(k+1)^2-1$ of degree $T$, density $d$ and order at least $q^{2k}-1$}.

Each proposition is proven via explicit construction of $F(n)$.

\section{Remarks on the size of numerical parameters and choice of generators}

We propose a constructive method of generation of bijective families $F(n)$ of stable bijective multivariate maps of vector space
$F_q$ of dimension $n=k(k+1)$ or $n=(k+1)^2-1$ of prescribed polynomial degree $s$, polynomial density $d \ge 1/2$ and exponential order 
 $\ge q^k-1$. It allows to generate $F(n)$ and its inverse in polynomial time.

We suggest the described above algorithms with the usage of such $F(n)$ defined over $F_q$ and integer parameters $k_A$ and $k_B$ are of size 
$O(n^{d_A})$ and $O(n^{d_B})$. Such choice insures the polynomial time for the computation of ${F(n)}^{k_A}$, ${F(n)}^{k_B}$ and
${F(n)}^{k_Ak_B}$ in the case of key exchange protocol.
Notice that density of ${F(n)}^{K_A}$ can be higher than $d$, it is bounded from above by $s$.
Alice sends the generator $F(n)$, ${F(n)}^{k_A}$ in a standard form to Bob together with parameter $d_B$ which restricts his choice of the
private key.
 
The precise computation of the order of $F(n)$ is a difficult task.
In the case of El Gamal algorithm our scheme below allows Alice to compute ${F(n)}^{-1}$ without the knowledge of order of $F(n)$.

Shifted El Gamal cryptosystem requires additional transformation $h(n)$. Notice that the map ${F(n)}^{-1}$ is hidden.

The known function is $M(n)=h(n)F(n){h(n)}^{-1}$. The adversary has to solve for $k_B$ the discrete logarithm problem with the base $M(n)$
and given $M(n) ^{k_B}$. Our method allows to generate both $F(n)$ and $M(n)$ as stable bijective transformations with prescribed degrees
$n_F$ and $n_M$ and densities $d_F$ and $d_M$.

\section{Double Schubert graphs and automata for the generation of stable maps}

		We define Double Schubert Graph $DS(k, K)$ over commutative ring $K$ as incidence structure defined as disjoint union of
 points from $PS=\{ ({\rm x})=(x_1, x_2, \dots, x_k, x_{1,1}, x_{1,2}, \dots x_{k,k})| \in ({\rm x})\in K^{(k+1)k} \}$ and
lines from $LS=\{[{\rm y}]=[y_1, y_2, \dots, y_k, y_{1,1}, y_{1,2}, \dots y_{k,k}]| \in ({\rm y})\in K^{(k+1)k} \}$ where
 $({\rm x})$ is incident to $[{\rm y}]$ if and only if $x_{i,j}-y_{i,j}=x_iy_j$ for $i=1, 2, \dots, k$, $j=1,2, \dots, k$.
It is convenient to assume that indices of kind $i,j$ are placed in lexicographical order.

REMARK. {\em The term {\em Double Schubert Graphs} is chosen because points and lines
of $DS(k, F_q)$ can be treated as subspaces of ${F_q}^{2k+1}$ of dimensions $k+1$ and $k$ which form two largest Schubert cells. 
Recall that the largest Schubert cell is the largest orbit of group of unitriangular matrices acting on the variety of subsets of
given dimensions.} (see \cite{12} and further references).

We define the colour of point $({\rm x})=(x_1, x_2, \dots, x_k, x_{1,1}, x_{1,2}, \dots x_{k,k})$ from	$PS$ as tuple
$(x_1, x_2, \dots, x_k)$ and the colour of line 

$[{\rm y}]=[y_1, y_2, \dots, y_k, y_{1,1}, y_{1,2}, \dots y_{k,k}]$ as tuple
$(y_1, y_2, \dots, y_k)$. For each vertex ${\rm v}$ of $DS(k, K)$ there is a unique neighbour $N_{\alpha}({\rm v})$ of given colour
$\alpha=(a_1, a_2, \dots, a_k)$, $a_i \in K$, $i=1, 2, \dots, k$.

The {\rm symbolic colour} $g$ from $K[z_1, z_2, \dots, z_k]^k$ of kind	$f_1(z_1, z_2, \dots, z_k)$, $f_2(z_1, z_2, \dots, z_k)$, $\dots$, 
$f_k(z_1, z_2, \dots, z_k)$,
where $f_i$ are polynomials from $K[z_1, z_2, \dots, z_k]$ defines the 
 neighbouring line of point $({\rm x})$ with colour 

$(f_1(x_1, x_2,, \dots, x_k), f_2(x_1, x_2, \dots, x_k), \dots , 
f_k(x_1, x_2, \dots, x_k)$.

Let us consider a tuple of symbolic colours $(g_1, g_2, \dots, g_{2t})\in K[z_1, z_2, \dots, z_k]^{k}$ and
the map $F$ of
 $PS$ to itself which sends the point $({\rm x})$ to the end ${\rm v}_{2t}$ of the chain
 ${\rm v}_0$, ${\rm v}_1$, $\dots$, ${\rm v}_{2t}$, where $({\rm x})={\rm v}_0$, ${\rm v}_iI {\rm v}_{i+1}$, 
$i=0, 1, \dots, 2t-1$ and $\rho({\rm v}_j)=g_j(x_1, x_2, \dots, x_k)$, $j=1, 2, \dots, 2t$.
We refer to $F$ as closed point to point computation with the symbolic key  $(g_1, g_2, \dots, g_{2t})$.
As it follows from definitions $F=F_{g_1, g_2, \dots, g_{2t}}$ is a multivariate map of $K^{k(k+1)}$ to itself.
When symbolic key is given $F$ can be computed in a standard form via elementary operations of addition and multiplication
of the ring $K[x_1, x_2, \dots, x_k, x_{11}, x_{12}, \dots, x_{kk}]$. Recall that $(x_1, x_2, \dots, x_k, x_{11}, x_{12}, \dots, x_{kk})$ is our
plaintext treated as symbolic point of the graph.
Let
${\rm Sk}(k,K)$
 be the totality of all symbolic keys. We
  define product $(g_1, g_2, \dots, g_{2t})(h_1, h_2, \dots, h_{2s})$ of symbolic keys $(g_1, g_2, \dots, g_{2t})$ and
$({h_1, h_2, \dots, h_{2s}})$ as $({g_1, g_2, \dots, g_{2t}}, h_1(g_{2t}), h_2(g_{2t}), \dots, h_{2s}((g_{2t}))$.This product
converts ${\rm Sk}(k,K)$ to a semigroup.
It is easy to check that the map    $^k\eta$ sending 
$({g_1, g_2, \dots, g_{2t}})$ to  $F_{g_1, g_2, \dots, g_{2t}}$ is the homomorphism of ${\rm Sk}(k,K)$ onto
$C(K^n)$ where $n=k(k+1)$. We refer to $^k\eta$ as {\em linguistic retraction morphism}.

We write $fg$ to the composition $g(f({\rm x}))$. If ($g_1$, $g_2$, $\dots$ , $g_k$) 
are elements of affine Cremona group $C(K^k)$ 
 then
$F_{g_1, g_2, \dots, g_{2t}}= F_{g_1}F_{{g_1}^{-1}g_2}F_{{g_2}^{-1}g_3} \dots F_{g_{2t-1}^{-1}g_{2t}}$.

We refer for expression $F_{g_1, g_2, \dots, g_{2t}}$ as automaton presentation of $F$ with the symbolic key $g_1, g_2, \dots, g_{2t}$. 
Notice that if $g_{2t}$ is an element of affine Cremona group $C(K^k)$ then $F_{g_1, g_2, \dots, g_{2t}} \in C(K^{k(k+1)})$
and automaton presentation of its inverse is $F_{g_{2t}^{-1}g_{2t-1}, {g_{2t}}^{-1}g_{2t-2}, \dots, {g_{2t}}^{-1}g_{1}, {g_{2t}}^{-1}}$.

The restrictions on degrees and densities of multivariate maps $g_i$  of $K^k$ to $K^k$ and size of parameter $t$ allow to define
a polynomial map $F$ of polynomial degree and density. 

Let us assume that $g_i=({h_1}^i, {h_2}^i, \dots, {h_k}^i)$, $i=1,2, \dots, 2t$ is the symbolic key of the closed point to point computation $F=F(k)$ of the symbolic automaton $DS(k,K)$. We set that $g_0=({h_1}^0, {h_2}^0, \dots, {h_k}^0)=(x_1, x_2, \dots, x_k)$.
 We set that ${h_1}^0, {h_2}^0, \dots, {h_k}^0)=(z_1, z_2, \dots, z_k)$.
Then $F$ is a transformation of kind

$z_1 \rightarrow {h_1}^{2t}(z_1, z_2, \dots, z_k)$, $z_2 \rightarrow {h_2}^{2t}(z_1, z_2, \dots, z_k)$, $\dots$, $z_k \rightarrow {h_k}^{2t}(z_1, z_2, \dots, z_k))$

$z_{11} \rightarrow z_{11} - {h_1}^1z_1+{h_1}^1{h_1}^2-{h_1}^3{h_1}^2+{h_1}^3{h_1}^4 + \dots +{h_1}^{2t-1}{h_1}^{2t}$

$z_{12} \rightarrow z_{12}- {h_1}^1z_2+ {h_1}^1{h_2}^2-{h_1}^3{h_2}^2+{h_1}^3{h_1}^4 + \dots +{h_2}^{2t-1}{h_1}^{2t}$

$\dots$

$z_{kk} \rightarrow z_{kk}- {h_k}^1z_k+ {h_k}^1{h_k}^2-{h_k }^3{h_k}^2+{h_k}^3{h_k}^4 + \dots +{h_k}^{2t-1}{h_k}^{2t}$

LEMMA 1.

    {\em The degree of $F$ is bounded  by a maximum $M$ of 
$\gamma_{r,s,i}(n)= {\rm deg}({h_r}^i) + {\rm deg}({h_s}^{i+1})$, $0\le i \le 2t$, $1\le r \le k$, $1 \le s \le k$.
The density of $F$ is at most a  maximum of $d(r,s)$, where $d(r,s)-1$ is  the
sum of parameters ${\rm den}({h_r}^i)\times{\rm den}({h_s}^{i+1})$ for $i=0,1, \dots, 2t$}.

We say that closed point to point computation $F$ is balanced if its degree coincides with parameter $M$ of the previous lemma.
	
LEMMA 2.
 
{\em If the map $g_{2t}: K^k \rightarrow K^k$ is a bijection then the presentation defines
one to one transformation  of $PS=K^{k(k+1)}$ to itself. The order of $F$ is  bounded below by the order of $g_{2t}$}.

LEMMA 3.

{\em If the map $g_{2t}: K^k \rightarrow K^k$ is an affine bijective transformation (${\rm deg}( g_{2t})=1$) and the
computation  is balanced then the map $F$ is stable
one two one transformation  of $PS=K^{k(k+1)}$ to itself.}

PROOF OF THEOREM 1 and 2.

\begin{proof}

Theorem 1 and 2 can be deduced from lemmas 1,2 and 3. We assume that parameter $t$ is a constant and $n=(k+1)k$.
Let us choose $F$ as $F_{g_1, g_2, \dots, g_{2t}}$ such that $(g_{2t})\in AFL_n(K))$ and parameter $M$ of Lemma 1 equals $T$.
Other maps $g_i$, $1 \le i \le 2k-1$ can be chosen to keep the density of balanced $F$ in the interval $C_1n^d$ and $C_2n^d$ where $C_1$ and $C_2$ are constants, $C_1 \le C_2$.
In the case of Theorem 1 we can choose $g_{2t}$
as linear permutation map corresponding to cycle of length $k$. This parameter $k$ gives the lower bound for the order of bijective map $F$. 
In the case of theorem 2 we can take Singer cycle in ${F_q}^k$ as $g_{2t}$. So $|F| \ge g^k-1$.

\end{proof}

PROOF OF THEOREM 3.

\begin{proof}

Let us consider the edge of the graph $DS(2k, F_q)$. It contains a point $({\rm p})=(x_1, x_2, \dots, x_k, x_{11}, x_{12}, \dots, x_{kk})$
and incident line $[{\rm l}]$ of colour $(y_1, y_2, \dots, y_k)$. We consider a chain of kind $({\rm p})$, $[{\rm l}]$, 
$({\rm p}_1)$, $[{\rm l}_1]$, $({\rm p}_2)$, $[{\rm l}_2]$, $\dots$,  $({\rm p}_s)$, $[{\rm l}_s]$ of odd length $2s+1$ such that
$\rho({\rm p}_i)=g_i \in {F_q[x_1, x_2, \dots, x_k, y_1, y_2, \dots, y_k]}^k$, 
$\rho({\rm l}_i)=h_i \in {F_q[x_1, x_2, \dots, x_k, y_1, y_2, \dots, y_k]}^k$, $i=1, 2, \dots, s$.
If pair $(x_1, x_2, \dots, x_k) \rightarrow g_{2s}(x_1, x_2, \dots, x_k, y_1, y_2, \dots, y_k)$,
 $(y_1, y_2, \dots, y_k) \rightarrow h_{2s}(x_1, x_2,  \dots, x_k, y_1, y_2, \dots, y_k)$ defines bijective map $\Delta$
of ${F_q}^{2k}$, then the map $F$ sending edge $p, l$  to  the edge $p_{2s}, l_{2s}$ is bijective transformation of edge set $E(2k, q)$ of 
$DS(2k, F_q)$. Notice, that $E(2k, q)$ is isomorphic to ${F_q}^{k(k+1)+k}={F_q}^{(k+1)^2-1}$. If we chose $\Delta$ as Singer transformation of the vector space ${F_q}^{2k}$, then the order of $F$ will be bounded below by $q^{((k+1)^2-1)}-1$.
Similarly to lemma 1 the degree and density of $F$ are maximum of parameters ${\rm deg}(g_i) + {\rm deg}(h_i)$, $i=0, 1, 2, \dots, s$ ,
${\rm deg}(h_i) + {\rm deg}(g_{i+1})$, $i=0, 1, 2, \dots, s$  (${\rm deg}(g_0)={\rm deg}(h_0)=1$). So appropriate choice of the symbolic key
insure that degree of transformation  $F$ is $T$ and density $d$.

\end{proof}

\section{Double Schubert automaton as a stable groups generator}

We refer to a subgroup $G$ in $S(K^n)$ as a stable subgroup of degree $d$ if the maximal degree for its representative $g$ equals $d$. 

Let $AGS_n(K)$ be the semigroup of affine transformations of $K^n$, i. e. the group of all transformations of degree 1.

It is easy to see that symbolic keys of kind ($g_1$, $g_2$, $\dots$, $g_r$) of even length  from  the semigroup
 ${\rm Sk}(k,K)$ with $g_i\in AGS_k(K)$, $i={1,2,\dots, r-1}$ and
 $g_r\in AGL_k(K)$  form a subgroup. We denote it as ${\rm Lk}(k,K)$. The degree
of the transformation 
$F_{{g_1}, {g_2}, \dots, {g_r}}$ for $<g_1$, $g_2$, $\dots$, $g_r>$ from ${\rm Lk}(k,K)$ is bounded by 2.
  Let us consider the group $E_k(K)=^k \eta({\rm Lk}(k,K))$.
  As it follows from
lemma 1 group $E_k(K)$ is stable subgroup of degree 2 in $C(K^{n(n+1)})$. 
The family of groups $E_k(K)$ can be used  for the following cryptosystem which can process rather large file.
It consists on following protocol, step of exchange of encryption rules and encryption process.

PROTOCOL.

Correspondents use family of group $E_k(F_q)$ for chosen papameters $k$ and $q$.
Alice computes $n=k(k+1)$ and selects affine transformation  $T$ from $AGL_n(F_q)$. She computes
$T^{-1}$. Alice selects positive integers $t$  and $r$ together with two strings ${\rm a}=(g_1, g_2, \dots, g_k)$
and ${\rm b}={h_1, h_2, \dots, h_r}$ of elements $g_i$ and $h_j$ from $AGS_k(F_q)$ 
such that $g_t$ and $h_r$ are Singer cycles from $GL_k(F_q)$, i. e. elements of  order $q^k-1$.  

She uses the homomotphism  $\eta=^k\eta$ from the semigroup ${\rm Sk}(k,K)$ onto affine Cremona group $C(K^n)$ and   
computes $\eta(a)$ and $\eta(b)$. Alice forms elements $G=T\eta(a)T^{-1}$  and $H=T\eta(a)T^{-1}$ of orders at most
$q^k-1$. In fact high orders of these elements are insured by the choice of linear transformations $g_t$ and $h_r$.

Alice sends to Bob the transformations $G$ and $H$ presented in their standard forms 
$x_i\rightarrow ^ig(x_1, x_2, \dots, x_n)$, $x_i \rightarrow ^ih(x_1, x_2, \dots, x_n)$, $i=1,2,\dots,n$ where
monomial terms of polynomials $^ig$ and $j^h$ are listed in the lexicographical order.
 
Secondly Alice selects positive constant integers $k_A <q^k-1$ and $r_A<q^k-1$. She computes
standard form $G_A= H^{r_A}G^{k_A}H^{-r_A}$ and sends it to Bob.  

In his turn Bob selects parameters   $k_B < q^k-1$ and $r_A < q^k-1$. He computes standard form of
$Z_B=H^{r_B}{G_A}^{k_B}H^{-r_B}$ and keeps it safely.

Secondly Bob form $G_B= H^{r_B}G^{k_B}H^{-r_B}$ and sends its standard form to Alice.
She computes $Z_A= H^{r_A}{G_B}^{k_A}H^{-r_A}$. 

Noteworthy that $Z=Z_A=Z_B$ is a collision element of the protocol.

In fact Alice and Bob share an element $Z$ from the stable group $Y(k, F_q)=T^k\eta({\rm Sk}(k, F_q))T^{-1}$  
of degree 2. 

STEP 2. ENCRYPTION TOOLS EXCHANGE.

Alice takes different from $T$ affine transformation $T'$ such 
that $TT' \ne T'T$.

She forms $G'=T\eta(a')T'^{-1}$ and $H'=T\eta(b')T'^{-1}$ of orders at least  $q^k-1$.
Correspondents again execute the protocol and elaborate another collision map $Z'$.

Secondly Alice (or Bob) selects  extra two elements $T_1$ and $T_2$ from  $AGL_n(F_q)$
such that elements of each pair selected from $\{ T, T', T_1, T_2  \}$ does not commute.
 She takes two strings $b$ and $c$ from ${\rm Sk}(k, F_q)$ of length $O(1)$, computes $P=T_1\eta(b){T_1}^{-1}$ and
  $Q=T_2\eta(b){T_2}^{-1}$ together with their inverses.
Finally Alice sends $Z+P$ and $Z'+Q$ to  Bob via open channel.
He restores $P$ and $Q$.

ASYMMETRIC ENCRYPTION PROCEDURE.

Before the start of exchange session  Alicia and Bob 
agree on the tuple of integers $(\alpha_1, \alpha_2, \dots, \alpha_s)$ of length $s=O(1)$
({\em password of the session}).

 Bob takes his plaintext ${\rm p}= (p_1, p_2,\dots, p_n)$ and applies transformation $P$ to it $\alpha_1$ times and
gets ${P}^{\alpha_1}({\rm p}= ^1{\rm p}$. In similar way he constructs
${Q}^{\alpha_2}(^1{\rm p})= ^2{\rm p}$. Bob gets $^3{\rm p}$ via the multiple application of $P$ to $^2{\rm p}$.
Let us assume for simplicity that $s$ is even. Then continuation of the process of recurrent  applications of $P$ and $Q$
forms the output $P^{\alpha_1}Q^{\alpha_2}\dots P^{\alpha_s}({\rm p})={\rm y}$. So Bob sends the ciphertext ${\rm y}$ 
to his partner.

  Alice uses natural decryption procedure. She takes reverse word $(\alpha_s, \alpha_{s-1}, \dots, \alpha_1)$.
She applies $P^{-1}$ to the ciphertext ${\rm y}$ with multiplicity ${\alpha_s}$ and gets $^1{\rm y}$, applies $Q^{-1}$ to
$^1{\rm y}$, $\dots$. Continuation of this process gives her the plaintext ${\rm p}$.

 COMPLEXITY ESTIMATES.

Straightforward computation of the number of elementary operations shows that Alice can construct
multivariate map $G$ (as well as $H$, $G'$, $H'$ in $O(k^7)=O(n^{3.5})$. This bound can be used for time evaluation of 
 Step 2.

The execution time of presented above key agreement protocol is determined by the hardest operation to compute the composition
of two quadratic maps of dimension $n=k(k+1)$ given in their standard forms. This operation requires  $O(n^5)$. 

It is easy to see that encryption of single message costs standard for Multivariate Cryptography time $O(n^3)$

ELEMENTS OF CRYPTANALISIS.

In the case of agstract finite group $X$ twisted key agreement protocol with input elements $G\in X$, $H \in X$ and output
$Z\in X$ is known instrument of NONCOMMUTATIVE CRYPTOGRAPHY (see \cite{27}-\cite{38}). 
It based on complexity of Power Conjugacy Problem. Adversary can intercept  $G_A= H^{r_A}G^{k_A}H^{-r_A}$ (or  
$G_B= H^{r_B}G^{k_B}H^{-r_B}$ ). To break the protocol he $/$ she hast find the presentation of $G_A$ in the form of word of kind
$H^xG^yG^{-x}$ 

Currently algorithm with the joint usage of Turing machine and
  Quantum computer for breaking this problem  in the case of affine Cremona group $X=C(K^n)$ are unknown.
	So adversary has no chances to break the protocol. During single session of exchanges with the string
	$(\alpha_1, \alpha_2, \dots , \alpha_s)$
	adversary can estimate the degree of encryption
	map $U=P^{\alpha_1}Q^{\alpha_2}\dots P^{\alpha_s}$ as $D=2^d$, $d=\alpha_1+\alpha_2+ \dots + \alpha_s$.
	He$/$she can execute interception of $n^D$ pairs plaintext-ciphertext and try to approximate $U$ via costly linearisation attack.
	( the cost is at least $\ge O(n^{2D+1})$.
	To prevent this option correspondents can agree on the maximal number $M=cn^{D-1}$ of messages during the session.
	They can start new session with other selected string without repetition of the protocol.
	
	Noteworthy that increase of $s$ to $O(log_2(k))$ makes linearisation attacks unfeasible.
	So correspondents can work with unlimited session for which encryption costs $O(n^3log_2(n^{1/2}))$.
	
	REMARK 1. {\em Alice can increase the number of protocols sessions  from 2 to chosen $l$, $l \ge 2$. It gives her opportunity of safe 
	delivery of noncomputing transformations $G_1$, $G_2$, $\dots$, $G_l$ and use this larger set of generators in the described above encryption algorithm}.
	
	REMARK 2. {\em Alice can take $l=4$ and $G_3={G_1}^{-1}$ and $G_4={G_2}^{-1}$. In this case both sides have option to decrypt. So we have
	symmetric encryption algorithm.}

		\section{Conclusion}
		
	       Algebraic system on $K[x_1, x_2, \dots , x_n]$, where $K$ is a commutative ring with operations of addition, 
	multiplication and composition is the core part of Computer Algebra. Let  ${\rm deg}(f)$ be the degree of polynomial
	$f \in  K[x_1, x_2, \dots,  x_n]$, then ${\rm deg}(f)+{\rm deg}(g)={\rm max}({\rm deg}(f), {\rm deg}(g))$. The general formula for
	${\rm deg}(f(g))$ does not exist, only inequality ${\rm deg}(f(g)) \le {\rm deg}(f){\rm deg}(g)$ holds. 
	The addition and multiplication of $n$ polynomials from $K[x_1, x_2,\dots , x_n]$ of bounded degree can be computed in polynomial 
	time but there is no polynomial algorithm for the execution of the computation  of $n$ elements from 
	$K[x_1, x_2, \dots , x_n]$.
	    It means that in Cremona semigroup $S(K^n)$  of all endomorphisms of $K[x_1, x_2, \dots , x_n]$  the computation 
		of the product of $n$ representatives is 
			unfeasible task. Noteworthy that each endomorphism $F \in S(K^n)$ is defined by its values $f_i$ on $x_i$ and can be
			identified with the rule $x_i \rightarrow f_i(x_1, x_2, \dots, x_n)$, $i=1,2, \dots, n$, where $f_i$ is given via the list of its monomial terms written in the lexicographical order.
Noteworthy that the semigroup $S(K^n)$ and its subgroup $C(K^n)$ of all automorphisms of 
$K[x_1, x_2, \dots , x_n]$  are core objects of Multivariate Cryptography (MC). Classical Multivariate Cryptography considers only compositions of kind $T_1FT_2$ of single nonlinear element $F$ of small degree (2 or 3) with linear bijective endomorphisms $T_1$ and 
$T_2$ of degree 1 because of the heavy complexity for the computation of compositions.
  
	Discovery of large stable subsemigroups $X$ of $S(K^n)$ of degree bounded by constant degree $d$ gives new option.
One can compute the composition of $n$ representatives of $X$ in polynomial time. So Diffie-Hellman protocol or its modifications
with generator from $X$ are  possible. Security of them requires further investigations. The cases $d=2,3$ has computational advantage because the composition of two nonlinear map can be
computed in time $O(n^5)$ and $O(n^{13})$.  The existence of implemented models of quantum computers even with restricted number of cubits
stimulates studies of analogs of Diffie -Hellman protocol with at least two generators $g_1, g_2, \dots, g_s$ and
noncommutative semigroup $X=<g_1, g_2, \dots, g_s>$. Current paper contains description of such protocol in the case of stable subgroups
 $X$
of degree 2 in $S(K^n)$.  The security of this algorithms rests on hard Power Conjugacy Problem. In the case $K=F_q$ one can select generators of exponential order.
 For the construction of Postquantum Secure Cryptosystem we combine this protocol with asymmetrical encryption algorithm , which allows execution of encryption for Bob and decryption for Alice in times $O(n^3)$ and $O(n^2)$ respectively.  We consider the way to convert encryption procedure into symmetrical algorithm in previous section.

  Other cryptosystems with the same platform ot its expansion are presented in \cite{39}, \cite{40}. They use Word Decomposition Problem instead Power Conjugacy Problem. The implementations of such protocol of Noncommutative Cryptography for stable subgroups $X$ of degree 3 is
	described in \cite{24}.






\end{document}